\documentclass[traditabstract]{swsc} 

\usepackage{graphicx}
\usepackage{txfonts}
\usepackage{caption}
\usepackage{subcaption}
\usepackage{placeins}
\usepackage{epstopdf}
\usepackage[displaymath,mathlines]{lineno}
\usepackage[authoryear,round]{natbib}
\usepackage[backref]{hyperref}
\usepackage{url}
\usepackage{xcolor}

\hypersetup{colorlinks=true,citecolor=cyan,urlcolor=cyan,linkcolor=blue}

\begin{document}

   \title{The mini-neutron monitor: A new approach in neutron monitor design}

   \titlerunning{Mini-neutron monitor}

   \authorrunning{Strauss et al.}

   \author{Du Toit Strauss\inst{1}\thanks{Corresponding author: \email{\href{mailto:dutoit.straussn@nwu.ac.za}{dutoit.strauss@nwu.ac.za}}}, Stepan Poluianov\inst{2,3}, Cobus van der Merwe\inst{1}, Hendrik Kr\"uger\inst{1}, Corrie Diedericks\inst{1}, Helena Kr\"uger\inst{1}, Ilya Usoskin\inst{2,3}, Bernd Heber\inst{4}, Rendani Nndanganeni\inst{5}, Juanjo Blanco-\'Avalos\inst{6}, Ignacio Garc\'ia-Tejedor\inst{6}, Konstantin Herbst\inst{4}, Rogelio Caballero-Lopez\inst{7}, Katlego Moloto\inst{1}, Alejandro Lara\inst{7}, Michael Walter\inst{8}, Nigussie Mezgebe Giday\inst{9} \and  Rita Traversi\inst{10}}

   \institute{Center for Space Research, North-West University,
              Potchefstroom, South Africa
              \email{\href{mailto:dutoit.straussn@nwu.ac.za}{dutoit.strauss@nwu.ac.za}}
         \and 
             Space Physics and Astronomy Research Unit, University of Oulu, Finland
        \and
             Sodankyl\"a Geophysical Observatory, University of Oulu, Finland
        \and
        Institut f\"ur Experimentelle und Angewandte Physik, Universit\"at Kiel, Kiel, Germany
                \and
        South African National Space Agency, Hermanus, South Africa
        \and
        Universidad de Alcal\'a (UAH), Dpto. Física y Matem\'aticas and Dpto. Autom\'atica, Campus Cient\'ifico-Tecnoló\'ogico (Externo), Alcal\'a de Henares, Madrid, Spain
        \and
        Instituto de Geof\'isica, Universidad Nacional Aut\'onoma de M\'exico, M\'exico D.F., M\'exico
        \and
        Deutsches Elektronen-Synchrotron DESY in Zeuthen, Germany
        \and
        Department of Space Science and Applications Research, Ethiopian Space Science and Technology Institute, Addis Ababa, Ethiopia
        \and
        Department of Chemistry ``Ugo Schiff", University of Florence, Italy
             }

 
  \abstract
   {The near-Earth cosmic ray flux has been monitored for more than 70 years by a network of ground-based neutron monitors (NMs). With the ever-increasing importance of quantifying the radiation risk and effects of cosmic rays for, e.g., air and space-travel, it is essential to continue operating the existing NM stations, while expanding this crucial network. In this paper, we discuss a smaller and cost-effective version of the traditional NM, the mini-NM. These monitors can be deployed with ease, even to extremely remote locations, where they operate in a semi-autonomous fashion. We believe that the mini-NM, therefore, offers the opportunity to increase the sensitivity and expand the coverage of the existing NM network, making this network more suitable to near-real-time monitoring for space weather applications. In this paper, we present the technical details of the mini-NM's design and operation, and present a summary of the initial tests and science results.  }


   \maketitle

\section{Introduction}

{Cosmic rays, including sporadic solar energetic particle (SEP) events, are the dominant source} of hazardous {atmospheric} radiation at, and above, aviation altitudes, making the long-term monitoring of cosmic ray levels essential. The latter is especially important since the International Civil Aviation Organization's (ICAO's)\footnote{ICAO State Letter, Proposals for the amendment of Annex 3 and consequential Amendments to Annex 15, PANS‐ABC and PANS‐ATM, AN 10/1‐17/41, 7 April 2017.} recent amendments to the meteorological procedures to include space weather warnings and/or predictions as part of flight planning. A reliable near real-time estimate of the cosmic ray flux is therefore essential. Moreover, ICAO requires ground-based cosmic ray observations that can serve as a proxy for the levels of radiation at aviation altitudes, making the {\it neutron monitor} (NM) {a reference space weather instrument for aviation dosimetry studies, but also more generally for monitoring the levels of cosmic radiation that may cause unwanted effects, including single-event upsets \citep[see e.g.][]{TaberNormand1993,Dyeretal2007,Dyeretal2009}, in avionics and other sensitive electronics devices (see also the IEC 62396-1:2016 standard\footnote{Available at \url{https://webstore.iec.ch/publication/24053}.}).}\\

For more than 70 years, NMs have been the most cost-effective instrument to monitor the levels of cosmic radiation near Earth \citep[see the recent review by][]{Butikofer2018}. These instruments, however, do not measure the cosmic radiation directly, but rather the flux of secondary and tertiary neutrons produced in the atmosphere, and in the detector itself. The count rate of an NM can thus be written, in the most basic formulation, as

\begin{equation}
    N(P_c,t) = \sum_i \int_{P_c}^{\infty} j_i (P,t) Y_i (P,t, \ldots) dP,
\end{equation}

where $P_c$ is the cutoff rigidity (minimum rigidity particle that can reach the detector), $i$ represents the particle distribution under consideration {(e.g. protons and heavier nuclei),} and $Y_i (P,t, \ldots)$ is the so-called yield function that represents the response of the instrument on the unit flux of primary cosmic rays with rigidity $P$, including atmospheric and instrumental effects. The latter therefore depends on various parameters that are unique to each monitor \citep[see e.g.][]{rogelio2016,clemdorman2000}. In order to reconstruct the primary cosmic ray flux, $j_i(P,t)$, the yield function must be known for a number of stations characterized by different $P_c$ values \citep[e.g.][]{mishevetal2020}. {Once the atmospheric yield functions are known, continuous spectral information about cosmic ray modulation (primary spectra above the atmosphere and magnetosphere) can be obtained from neutron monitor differential response functions. These response functions can be derived from the count rates of the world-wide neutron monitor network if individual NMs are inter-calibrated \citep[e.g.][]{moraal2001}.}\\

To inter-calibrate the world-wide network of NMs, \citet{moraal2001} initiated the development of a smaller and lighter version of the standard neutron monitor, a {\it calibration NM}. The aim was to design a portable NM that can easily be shipped and installed, even at very remote locations. Although this calibration process was mostly successful \citep[e.g.][]{kruger2003}, it was quickly realized that, under the right circumstances, a calibration NM's data can be of sufficient quality to supplement the measurements of traditional NMs. It was subsequently proposed by \citet{kruger2015} to refer to these smaller monitors as {\it mini-NMs}, with mini-NMs installed at the Concordia \citep[][]{stepan2015,usoskin2015} and Neumayer III \citep[][]{heber2015} research stations in Antarctica. Since then, the mini-NM's design has been constantly optimized and the counting electronics updated. In this paper, we describe the most recent version of the mini-NM, including mechanical design, electronics systems, and software implemented, along with initial tests of the data acquired.\\


\begin{figure}
\includegraphics[width=0.5\textwidth]{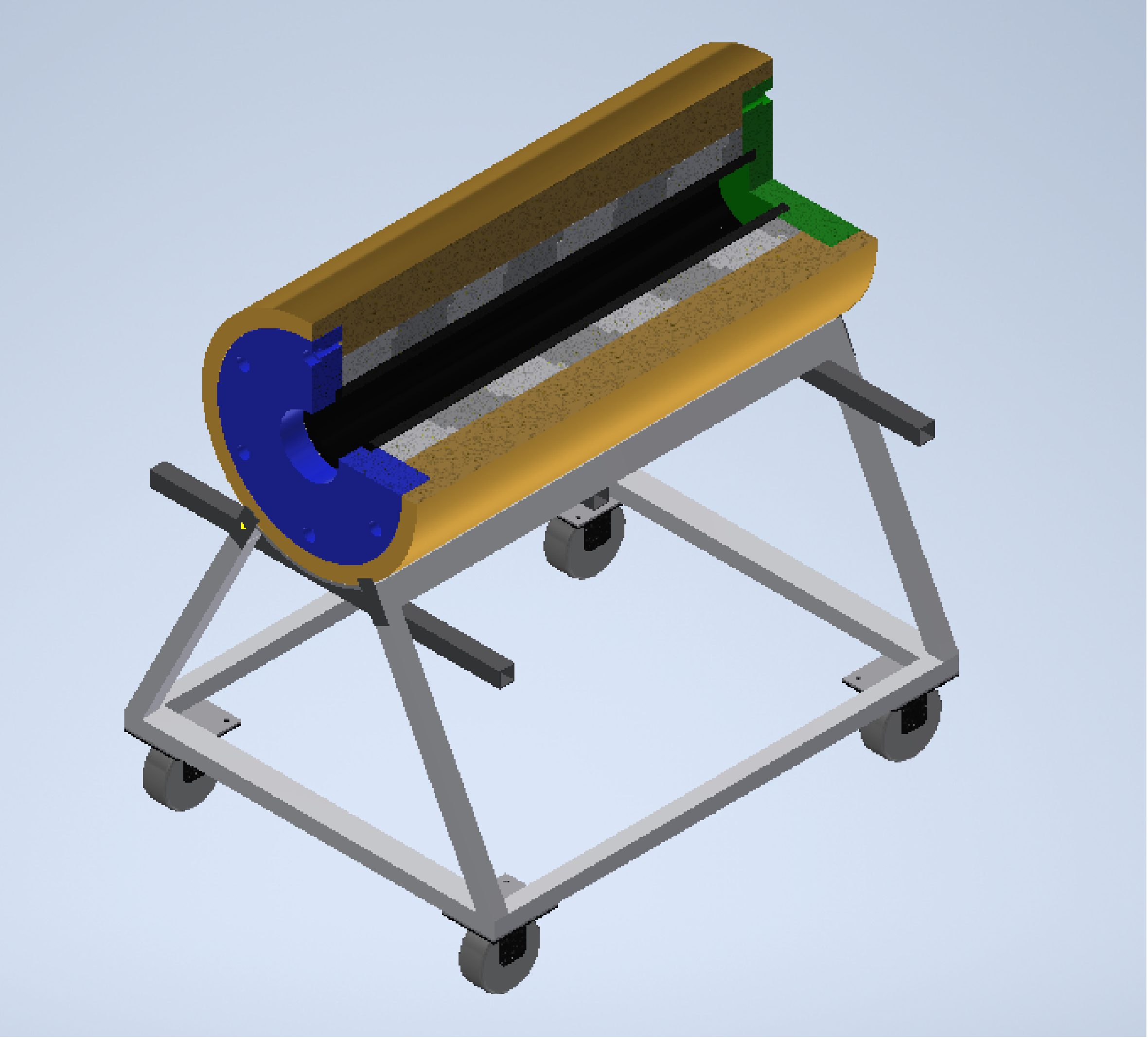}
\caption{A sketch of the standard mini-NM set-up. The reflector (yellow), moderator (black), front- and end-cap (blue and green), and lead rings (dark and light gray) are indicated.}
\label{fig:NM_sketch}
\end{figure}

\begin{figure*}
\includegraphics[width=\textwidth]{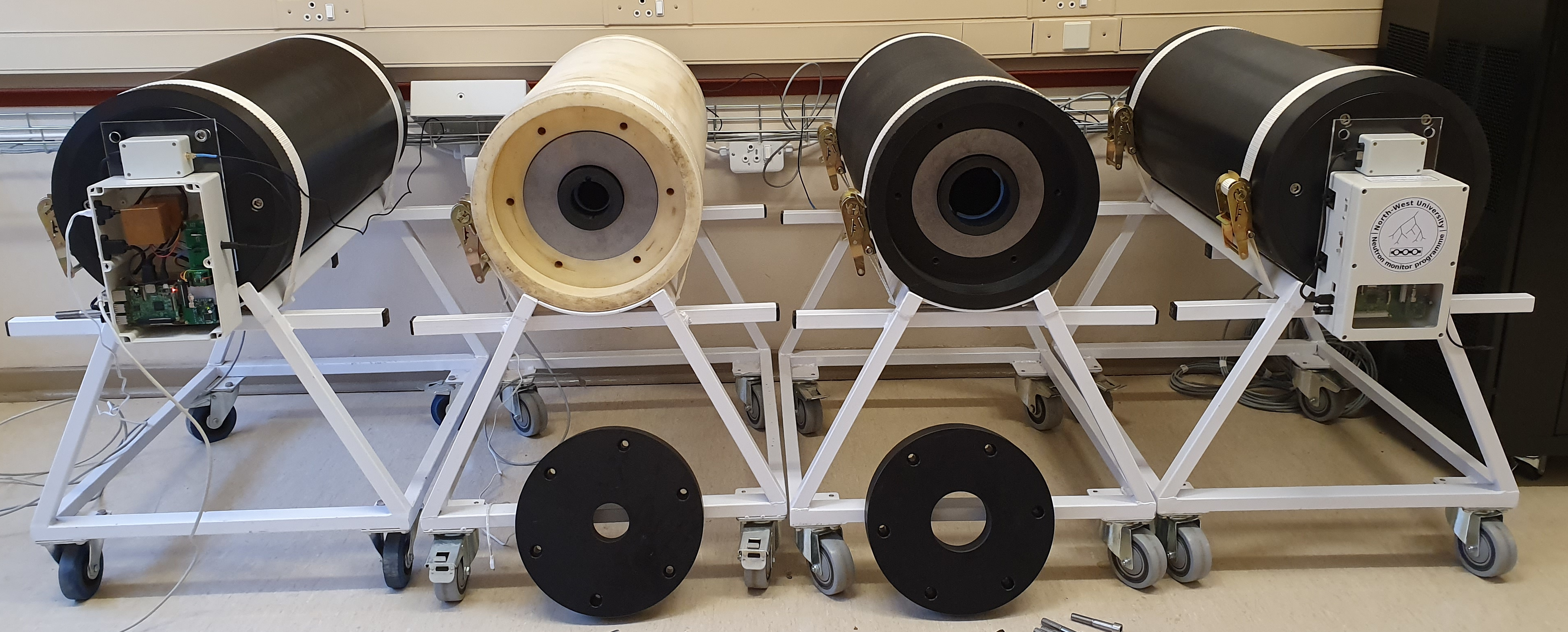}
\caption{Mini-NMs undergoing testing. The front covers were removed on two monitors to show the lead rings and the internal moderators.}
\label{fig:NM_testing}
\end{figure*} 

\section{Basic neutron monitor design}

The design of the basic NM has changed very little  over the years since its first development \citep[see the review by][]{simpson2000}, until the most recent implementation \citep[e.g.][]{signoretti2011,medina2013,jung2016}, consisting of an outer reflector (shielding the instrument from external thermal neutrons, while containing neutrons produced within the NM), a lead producer (producing additional {low energy} neutrons via nuclear interactions to increase the monitor's detection efficiency), and an inner moderator (thermalizing the newly produced neutrons). The thermal neutrons are then counted via pulses produced in a proportional counter tube, traditionally filled with either BF$_3$ or $^3$He gas. For both of these filling gases, an electrical pulse is produced via a neutron capture process \citep{knoll12010}, i.e.

\begin{eqnarray}
^{10}\mathrm{B} + \mathrm{n} & \rightarrow & \left\{ \begin{array}{ccc}
     ^{7}\mathrm{Li} + \alpha + 2.78 \, \mathrm{MeV} \\
     ^{7}\mathrm{Li}^{*} + \alpha + 2.30 \, \mathrm{MeV} \end{array} \right. 
\end{eqnarray}

or

\begin{equation}
    ^{3}\mathrm{He} + \mathrm{n} \rightarrow \,  ^{3}\mathrm{H} +\mathrm{p} + 0.76 \, \mathrm{MeV}
\end{equation}

that ionizes a small portion of the filling gas. Note that only a small fraction ($\sim 6\%$) of the reactions leaves $^7$Li in the ground state, releasing 2.78 MeV in kinetic energy, while the majority of the reactions leave $^7$Li in the first excited state, releasing only 2.3 MeV in kinetic energy. Some NMs, sometimes referred to as {\it neutron moderated detectors} (NMDs), do not have the lead producer and are sensitive to neutrons in a slightly lower energy range. These NMDs are referred to as ``bare", lead-free detectors, or even ``wax" NMs in certain texts where the reflector and moderator in older set-ups were manufactured from paraffin wax. {The physics of neutron detection in NMs and NMDs are identical, but the count rates of NMDs are usually lower, and the instrument responds to lower energy incident particles, due to the absence of a producer.}\\

\subsection{Mechanical design}

A sketch of the standard mini-NM design is shown in Figure \ref{fig:NM_sketch}. The outer reflector consists of the main cylinder (yellow) and front and end covers (blue and green), manufactured from 75mm thick rolled polyethylene sheets. Lead rings of a maximum diameter of 206mm are shown in light and dark gray, while the inner moderator (black) is also manufactured from rolled polyethylene sheets of thickness 15mm. The inner diameter of the lead rings, and the diameter of the moderator, is determined by the type of tube used in the monitor. For the purpose of this study, we have used three different tubes, with details summarized in Table \ref{Tab:tubes}. The completed mini-NM set-up, while undergoing testing, is shown in Figure \ref{fig:NM_testing}. {The dimensions of a completed mini-NM, with a trolley stands, are approximately 900mm (height) $\times$ 700mm (width) $\times$ 800mm (length) and the weight is approximately 250/100 kg with/without the lead rings.}

The volume of a mini-NM detector is notably smaller than that of a full-size standard NM64 (the 64 referring to the standardized NM set-up implemented in 1964 during the International Quiet Sun Year; IQSY). However, with a change in regulations of the ICAO, {related to the transport of dangerous goods}\footnote{\url{https://www.icao.int/safety/DangerousGoods/Pages/DGP23-Report.aspx}}, it became possible to significantly increase the allowed maximal pressure of the BF$_3$ tubes. The pressure in a mini-NM detector LND2043 is 930 mbar, while it is only 300 mbar in a standard NM64 detector. This partly compensates for the reduced efficiency of the mini-NM as compared to the much larger full-size NM64.\\

\begin{table*}
\caption{Comparison of the different tubes used for testing$^{\mathrm{a}}$.}             
\label{Tab:tubes}
\centering                          
\begin{tabular}{c c c c c c}        
\hline\hline
Tube naming &  & Gas & Outer diameter (mm) & Length (mm) & Recommended voltage (V) \\ 
\hline      
BF3 & LND2043 & BF$_3$ & 89 & 697 & 2120 -- 2370 \\
BF3(thin) & LND20366 & BF$_3$ & 51 & 652 & 900 -- 1150 \\
He3 & LND25373 & $^3$He & 51 & 652 & 1200 -- 1450 \\
\hline
\end{tabular}
\begin{list}{}{}
\item[$^{\mathrm{a}}$] Tubes are manufactured by LND Inc., USA and available through \url{https://www.lndinc.com/}.
\end{list}
\end{table*}

\begin{figure*}
\includegraphics[width=0.54\textwidth]{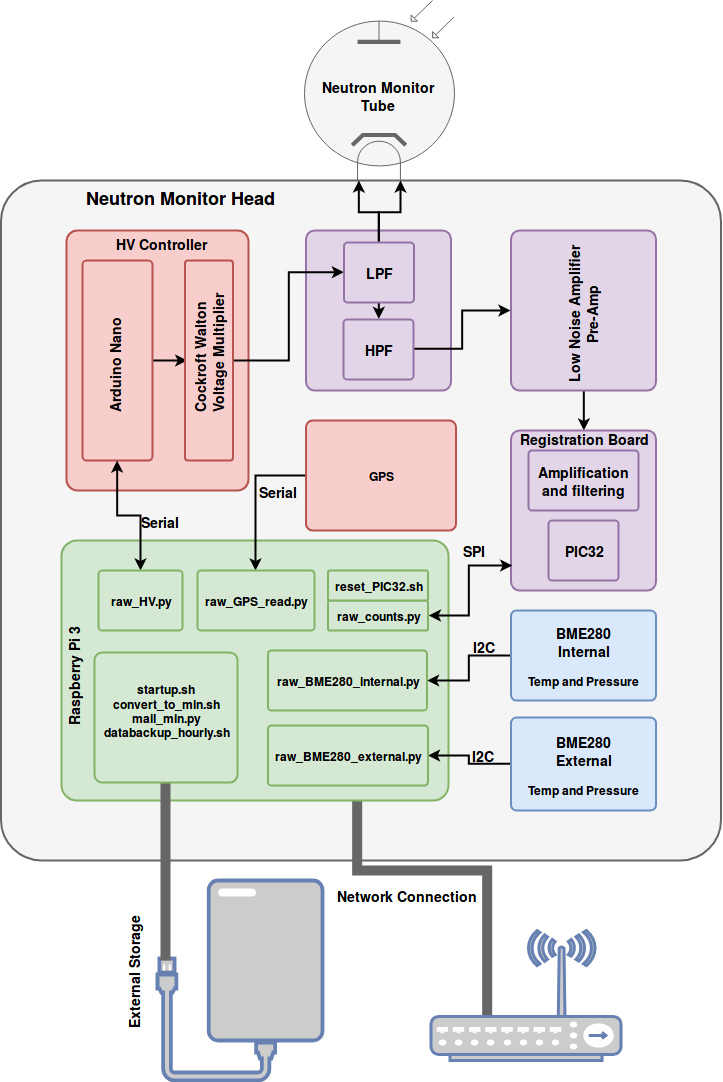}
\includegraphics[width=0.44\textwidth]{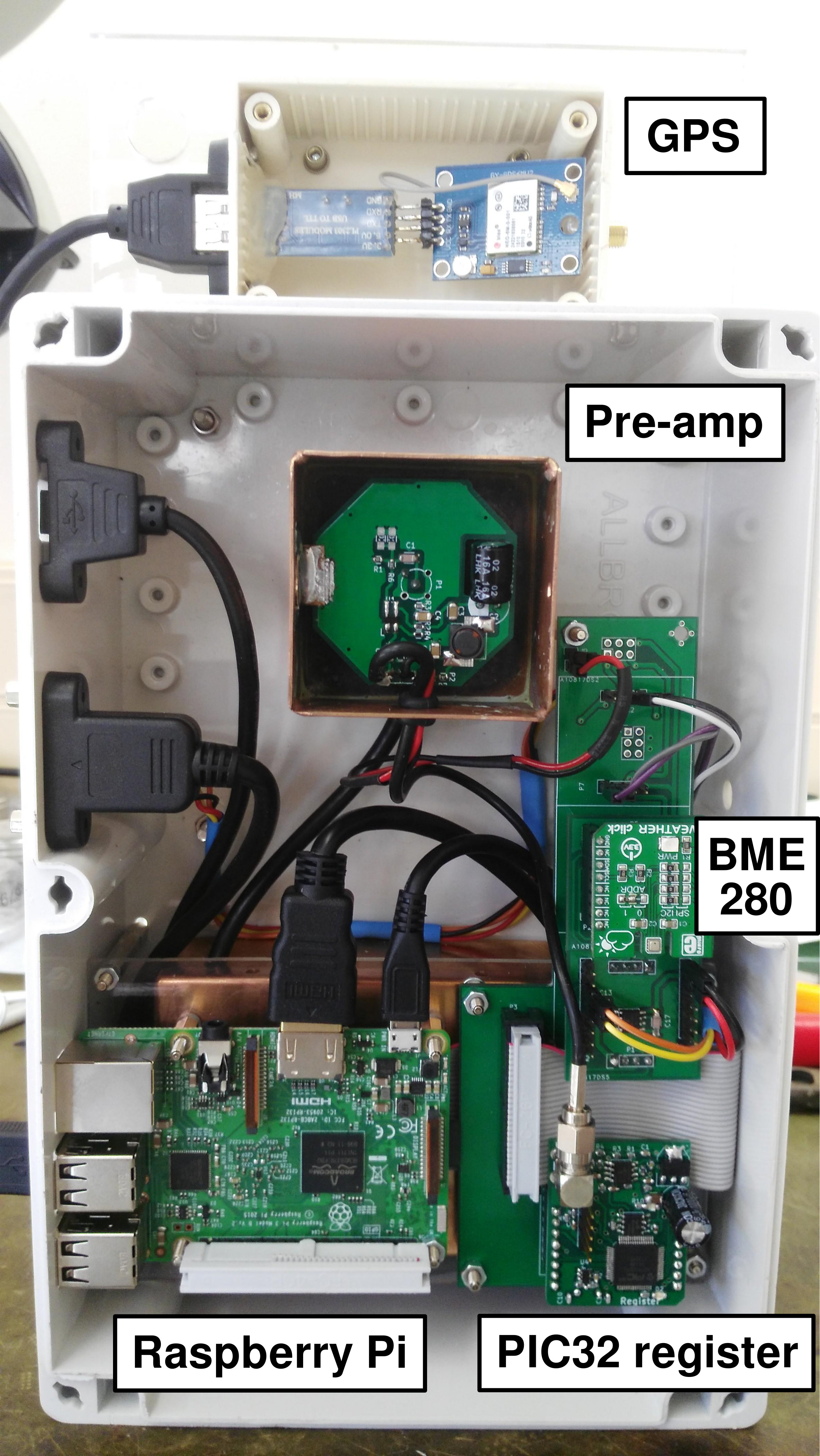}
\caption{The left panel shows a schematic representation of the mini-NM electronics head while the right panel shows the actual components, with the front cover removed.}
\label{fig:electronics}
\end{figure*}

\subsection{Electronics}

Although the mechanical design of a mini-NM, similar to a standard NM, is simple, the associated electronics are usually the most complicated part. In designing the new version of the mini-NM's electronics, it was decided to use, as much as possible, affordable, off-the-shelf components. In addition, it was decided to design the electronics in a modular fashion, where defective parts can be quickly (and cheaply) replaced. A sketch of the electronics head is shown in the left panel of Figure \ref{fig:electronics}, while the right panel shows a finished unit. The different components are briefly discussed below. 

\subsubsection{Raspberry Pi micro-computer}

A Raspberry Pi 3 B\footnote{\url{https://www.raspberrypi.org/}}, a fully-functional single-board computer with GNU/Linux operating system is used to interface with the system components. Not only does the Raspberry Pi offer a significant amount of computing power but it also simplifies a number of data processing tasks: The on-board computing power enables data processing and data validation in real-time. This allows functionality like digital filters and enables units to alert or automatically reset if an error occurs. The built in network interface allows for remote monitoring, administration, and data synchronisation. The USB (universal serial bus) interface allows for a local data backup when network connectivity is interrupted or unavailable. The weakness of the Raspberry Pi is the limited write cycles on the SD (secure digital) card. This can be overcome by using Linux shared memory for intermediate data recording and higher quality (e.g. industrial rated) SD cards.  

\subsubsection{High voltage supply}

The high voltage supply consists of a microprocessor and a set of voltage doubler circuits also known as a Cockroft-Walton multiplier. A PID (proportional integral derivative) controller adjusts a pulse width modulated input signal to regulate the output voltage. The desired voltage is sent to the microcontroller over a serial connection, and the voltage value is returned to the Raspberry Pi. The output range of the high voltage supply can be set between 0.8 -- 4 kV, where the maximum voltage is also dependent on the inductance of the transformer, which can also be adjusted. The 5 V signal from the microcontroller switches a transistor which supplies a flyback transformer before the multiplication stage. This high-voltage power supply design is feasible as the neutron monitor tube requires very little current. The only real concern is the long term voltage stability and minimal high-frequency noise, which is less than 1 mV on 4 kV.\\

\subsubsection{Pre-amplifier}

The pre-amplifier uses a low noise MOSFET (metal-oxide-semiconductor field-effect transistor) for its first amplification stage. A DC (direct current) blocking capacitor is used to filter the pulses from the high voltage on the neutron monitor cathode. The MOSFET's low gate current allows the filtered pulses to trigger the amplifier. Careful design considerations were made with regard to the input impedance and sensitivity of the pre-amplifier. The input impedance determines the shape of the amplified pulse, and over biasing will make the MOSFET too sensitive with the result that amplified noise will then dominate the later stages of the system. The pre-amplifier is a critical component, and variations between heads are most likely related to manufacturing tolerances in the MOSFET and its biasing components. Band-pass filters are used in the second stage of amplification to suppress any noise from the pre-amplifier and to shape the amplified pulses.\\

\subsubsection{Analogue-to-digital converter}

A dedicated PIC32 microcontroller is used to sample the incoming pulses, after amplification, with a 10-bit analog-to-digital converter (ADC). The pulses are sampled at 2 MHz and stored in the microcontroller's buffer. The sampling discriminator can be set in software, only triggering samples above the set value. This software filter allows for further elimination of noisy data. The Raspberry Pi software interfaces with the PIC32 via SPI (serial peripheral interface) and clears the sample buffer. The raw pulse data is stored and processed by the Raspberry Pi software. \\

\subsubsection{Peripherals}

The environmental conditions are measured with two Adafruit\footnote{\url{https://www.adafruit.com/product/2652}} BME280 sensor boards (the BME sensor itself is produced by Bosch.\footnote{\url{https://www.bosch-sensortec.com/products/environmental-sensors/gas-sensors-bme680/}}) The BME280 sensor has a maximum absolute offset of 1 mbar, and a relative error of 0.12 mbar. We have, however, performed long-term tests of this sensor and have found any possible offsets to be constant over time. The relative (noise) error is further decreased in the mini-NM by sampling the pressure every second and using {one-minute averages} for all calculations. We believe that the BME sensor provides a very affordable and accurate alternative to the more expensive barometers currently used. Data is sent to the Raspberry Pi via an I2C (inter-integrated circuit) connection. Internal (meant to be placed against the tube inside of the moderator) and external (placed inside the electronics enclosure) pressure/temperature sensors are available. A redundant pressure measurement can also help with the identification of any possible absolute calibration issues when installing a new pressure sensor or to test for possible drifts in a defective sensor. \\

The default time synchronization is realized with the Network Time Protocol (NTP), when the mini-NM receives time signals from an NTP server via a network connection. However, there is an optional GPS (global positioning system) unit that can be connected via a USB connector on the Raspberry Pi that sends location data and a signal for time synchronisation (when an NTP server is not available) over the USB-serial connection.

\subsection{Software}

Similar to the electronics, the software used on the mini-NM was designed to be modular with scripts written using the Python programming language \footnote{\url{https://www.python.org/}} or simple shell scripts. Specific scripts, therefore, have very specific purposes, e.g., a stand-alone Python script is used to read the raw GPS data over a serial connection and write the raw data to a file located on the Raspberry Pi. Similar scripts read data from the ADC, the BME sensors, and communicate with the Arduino nano \footnote{\url{https://www.arduino.cc/en/Guide/ArduinoNano}} which controls the HV. Once the raw data are collected in their native time resolution, the data is averaged into {one-minute time resolution}.\\

These scripts are continually being refined and updated, with the latest version available through a Creative Commons non-commercial license\footnote{\url{https://fskbhe1.puk.ac.za/neutronmonitor/Mini_NM/Software_images/}}.


\begin{figure*}
\begin{center}
\includegraphics[width=0.41\textwidth]{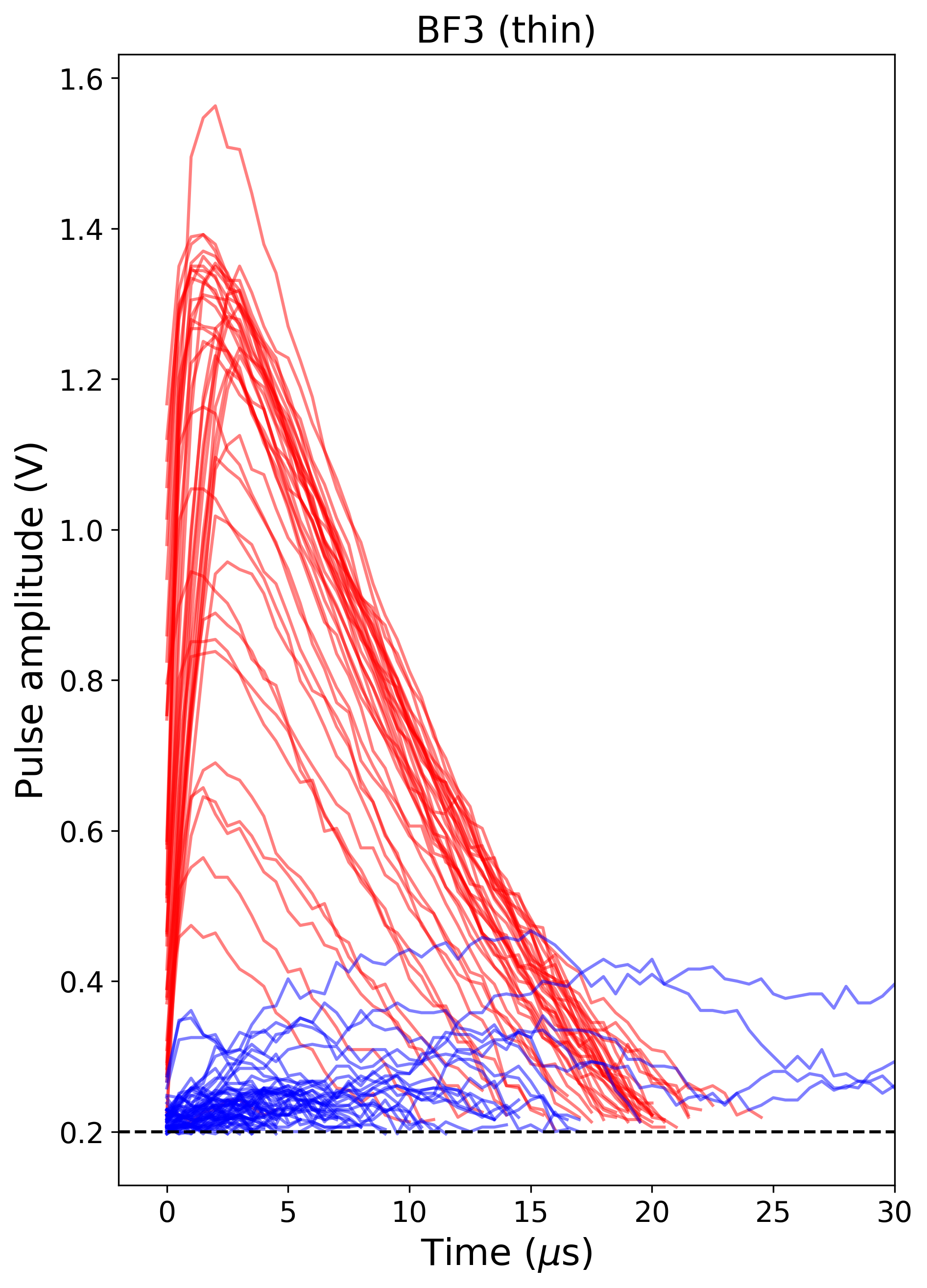}
\includegraphics[angle=00,width=0.58\textwidth]{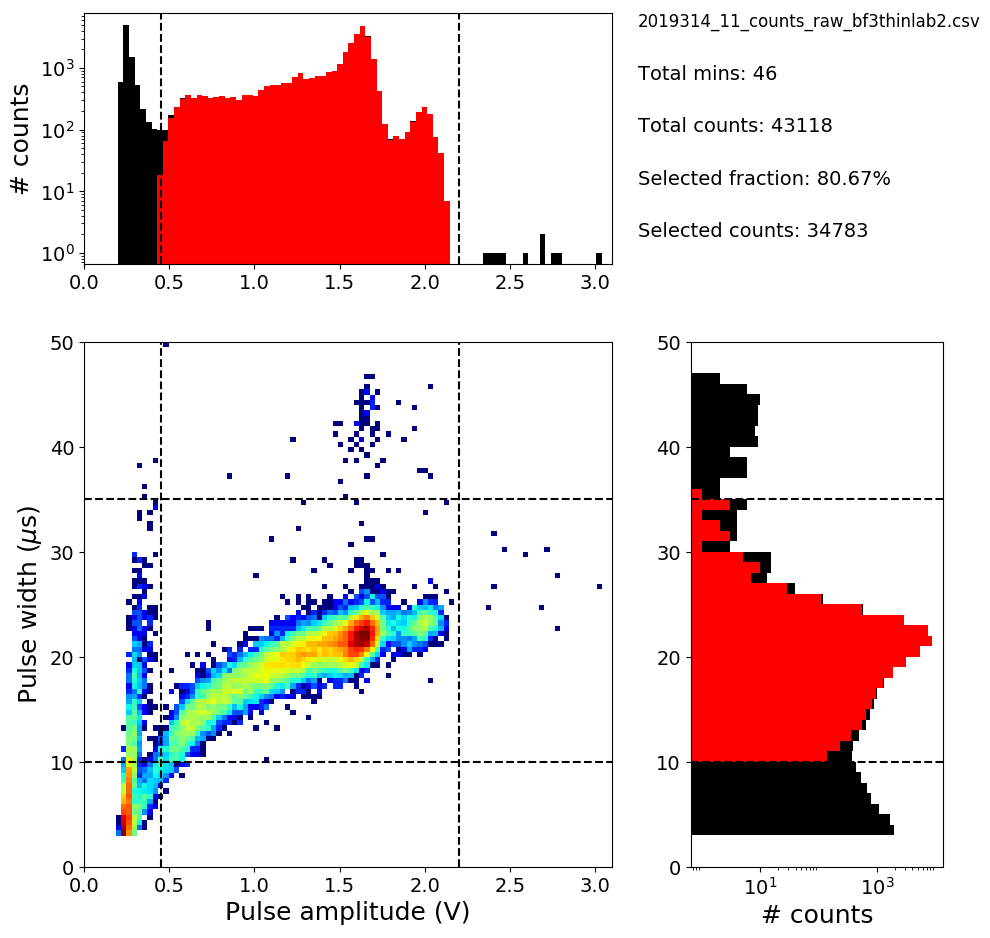}
\end{center}
\caption{Left panel: A selection of pulses registered from the thin BF$_3$ tube. Red pulses are assumed to be observations, while blue pulses indicate noise. Right panel: A 2D pulse width-amplitude histogram as registered from the same tube. The distribution is also projected onto each plane. The dashed lines show the cut applied to the pulses in order to discriminate between measurements and noise. On the projected histograms, all the pulses are indicated by a black histogram, while the red histogram indicated pulses assumed to be measurements free from noise.}
\label{Fig:pulses}
\end{figure*}

\begin{figure*}
\begin{center}
\includegraphics[width=0.49\textwidth]{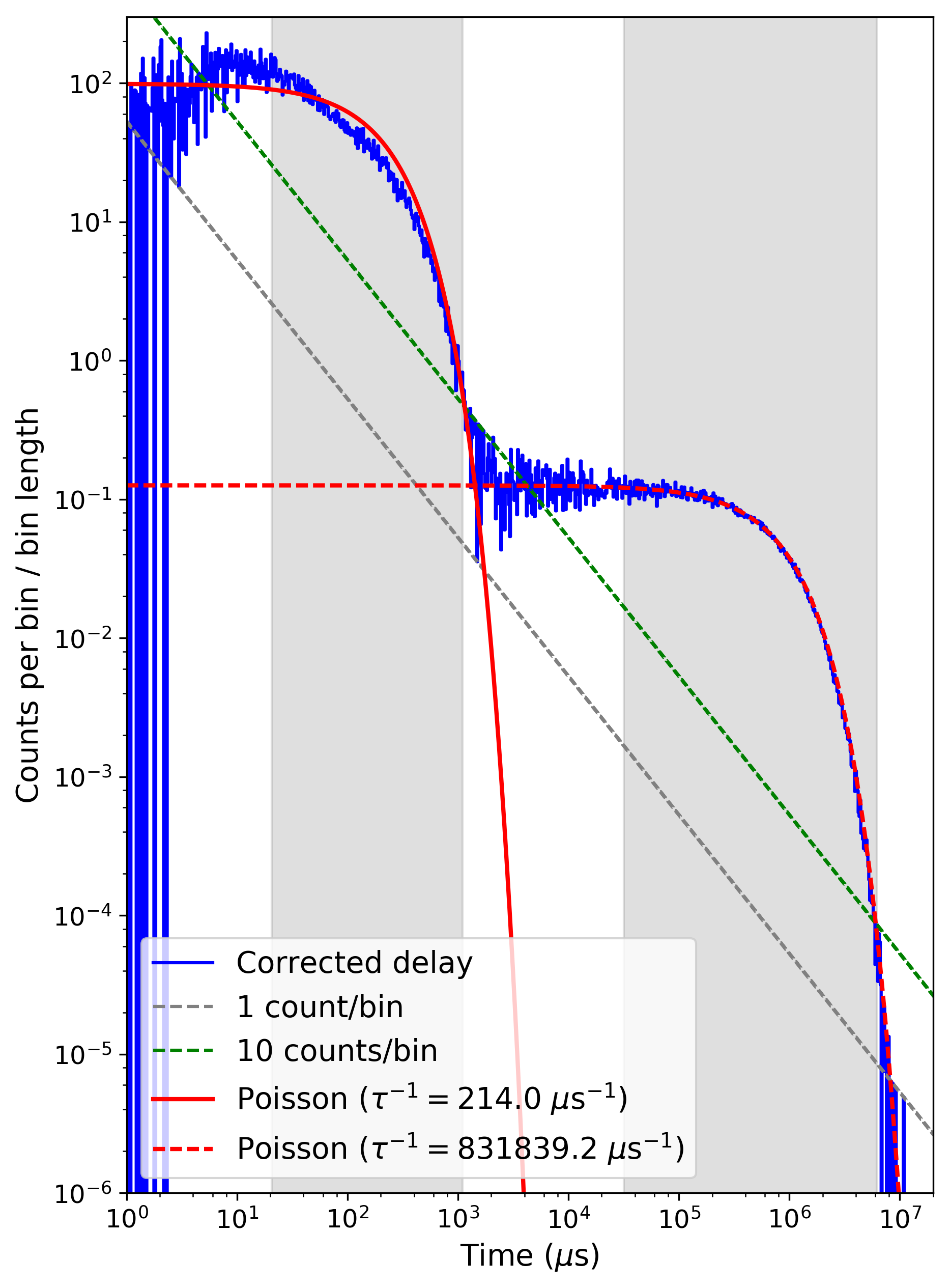}
\includegraphics[width=0.49\textwidth]{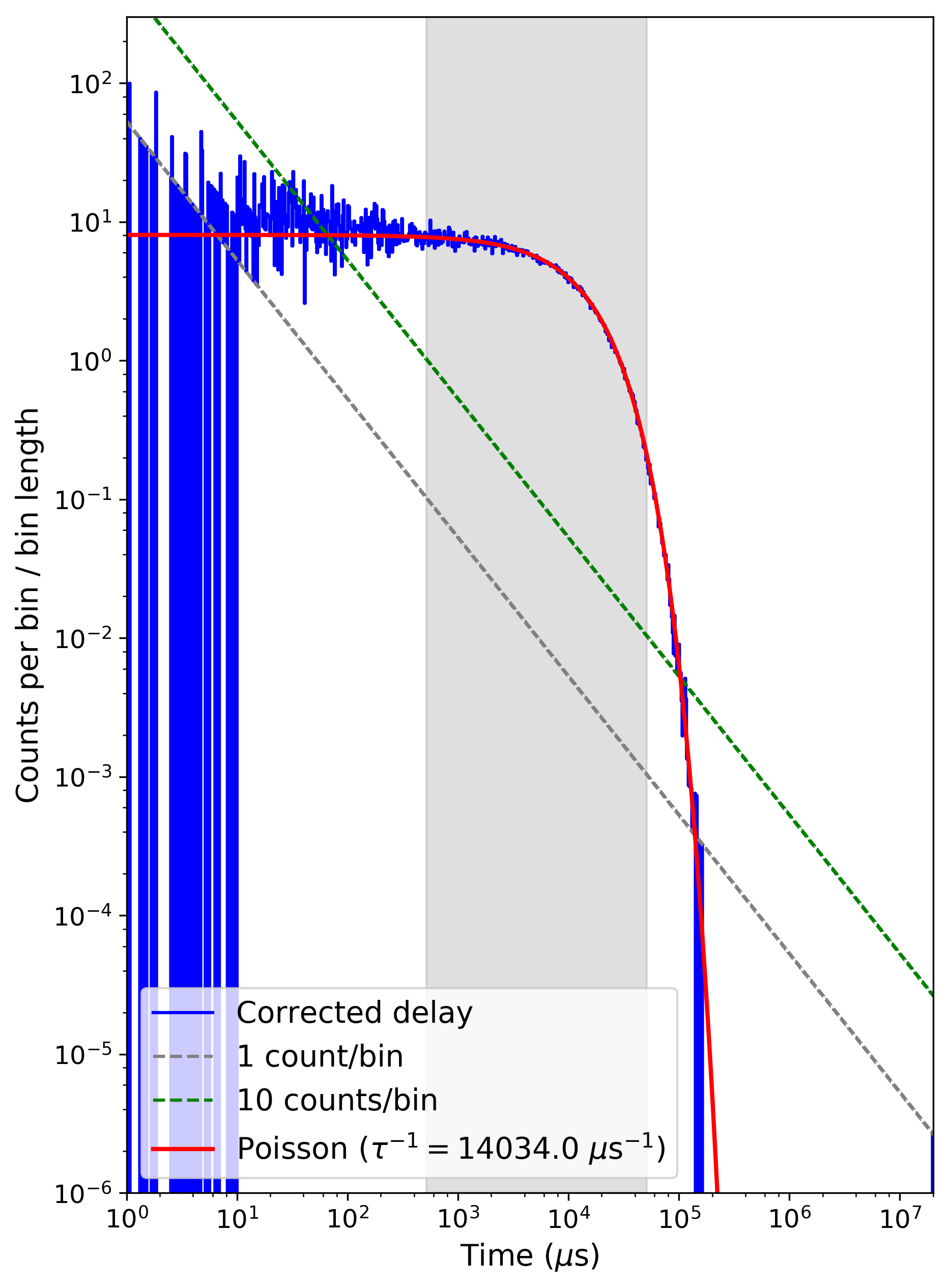}
\caption{The waiting time distribution (blue histogram) for cosmic ray neutrons as observed with the $^3$He tube (left panel) and for the test source (right panel). The red curves show fitted Poisson distributions, while the dashed gray and green lines indicate bins containing either 1 or 10 counts.}
\label{fig:multi}
\end{center}
\end{figure*}

\begin{figure*}
\begin{center}
\includegraphics[width=0.99\textwidth]{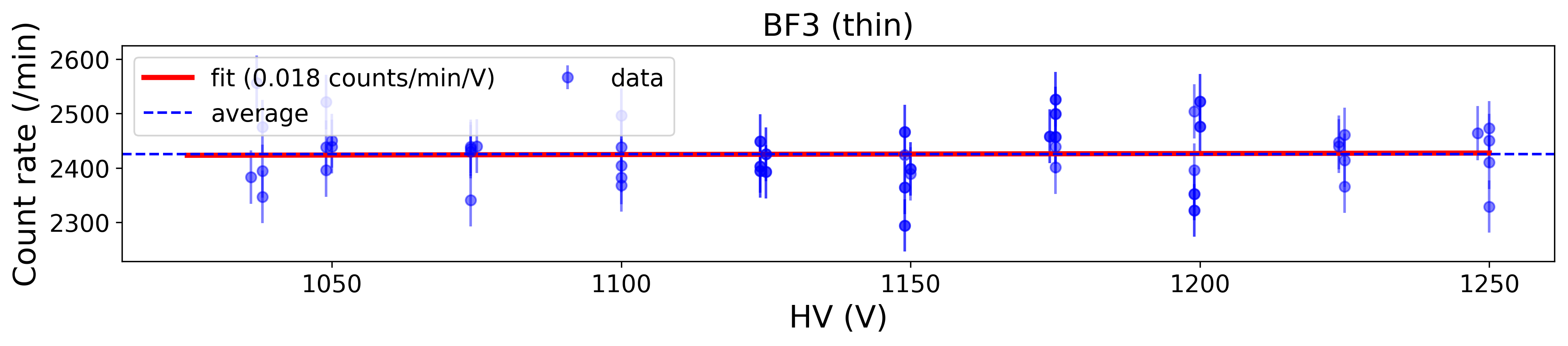}
\includegraphics[width=0.99\textwidth]{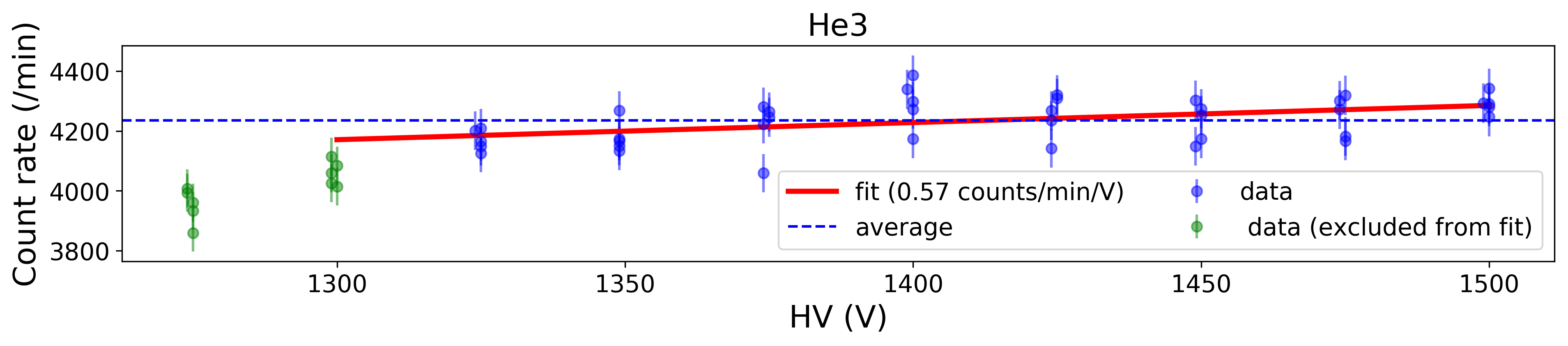}
\includegraphics[width=0.99\textwidth]{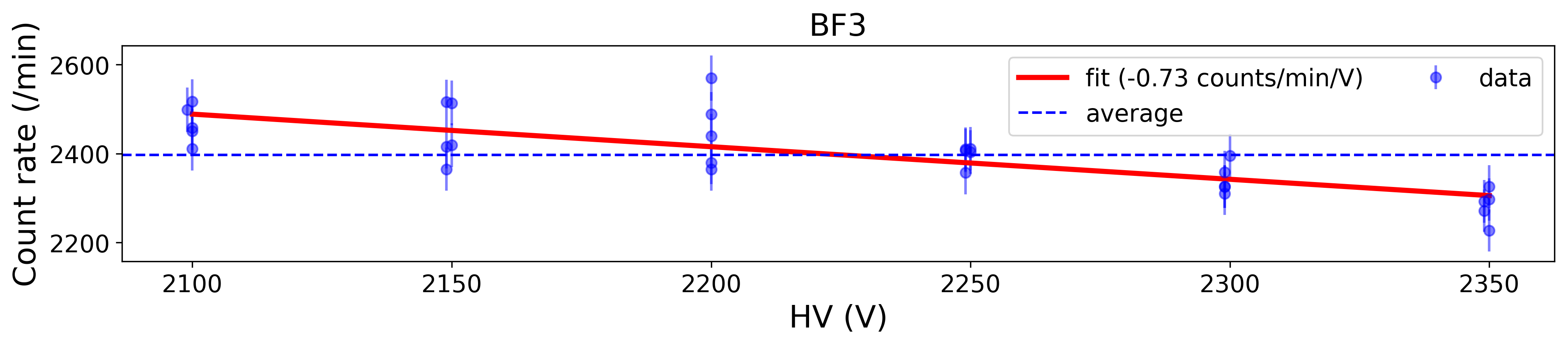}
\caption{The dependence of the observed count rate on variations of the HV, as measured for different tubes. The average count rate for each tube is indicated by the dashed line, while the red line shows the results of a linear regression. For the middle panel, green data points were excluded when performing the regression.}
\label{fig:HV_testing}
\end{center}
\end{figure*}

\section{Testing: Neutron source}

Initial testing of the mini-NM was performed with an $^{241}$Am-$^9$Be neutron source to improve measurement statistics, but also to eliminate any possible environmental dependences.\\ 

\subsection{Pulse height-width diagrams}

The left panel of Figure \ref{Fig:pulses} shows a selection of pulses read-out of the mini-NM with the thin BF$_3$ tube inserted. Based on the maximum pulse height and length (i.e., width) of the pulses, they are divided into counts (indicated by red) and noise (blue curves). The ADC threshold is indicated on this figure as the dashed line at 0.2 V. Once the ADC voltage crosses this threshold, the start of a pulse is registered, and the pulse is digitized until the voltage drops below this level. For each pulse, the amplitude (i.e., maximum pulse height) and length are recorded, and, as discussed in the next section, this is used to discriminate between data and noise.\\

The right panel of Figure \ref{Fig:pulses} shows a binned 2D histogram of the pulse width and pulse histogram, again for the thin BF$_3$ tube. The distributions are also projected onto each plane. From the 2D histogram, we can clearly identify low amplitude noise below $\sim 0.5$ V, consisting of short and longer pulses. \\

Two peaks are observed in the pulse width-amplitude plot, near 1.6 V and 2 V, which is presumably due to the decay into $^7$Li$^*$ and $^7$Li, respectively. Also note the continuous decreasing pulse amplitude towards lower pulse amplitudes, until the noise threshold is reached. This is, most likely, due to the {\it wall effect}, resulting in some fraction of energy being deposited into the wall of the detector \citep[see the discussion by e.g.][]{knoll12010}.\\

By calculating and monitoring these distributions on, e.g., an hourly basis, allows us to determine and track the accuracy of the detector: Any external noise (e.g., mechanical vibrations or high-voltage ripple currents) will change the shape of the distributions, allowing us to identify and correct for such events.\\

Visualizing the pulses in such a 2D plane also allows us to perform a detailed separation of the pulses into noise and/or data (counts) by applying an appropriate cut through the pulse amplitude/width plane. As an example, here we perform a simple square cut (indicated by the dashed lines) and count all the pulses in these regions as data. The projections show all the pulses as black histograms, and selected data as the red histograms. In the future, more complex separation algorithms can be applied to the pulses.\\

\subsection{Multiplicities}

The new hardware design allows us to also examine the waiting time, $\Delta t$, between pulses down to the low $\mu$s range. {For the purpose of this study, we define the waiting time simply as the time between pulses, i.e. the time that elapses from the end of one pulse (ADC drops below the threshold) and the onset of the next (ADC crosses the threshold).} The electronics sample at a constant rate of 2MHz, and, in principle, the detector dead-time is only determined by the length of the pulses observed. In a standard NM with the lead producer in place, these so-called {\it multiplicity} plots show a double-Poisson distribution. Using cosmic ray measurements from the $^3$He tube, Figure \ref{fig:multi} shows a distribution of the waiting time between pulses for a full day of observation of cosmic ray neutrons as the blue histogram in the left panel. The grey and green dashed lines indicate bins with either 1 or 10 counts (data below these lines are considered statistically insignificant). The shaded parts of the distribution is used to fit two separate Poisson distributions in the form

\begin{linenomath*}
\begin{equation}
    \mathcal{D}(\Delta t) = \mathcal{D}_0  \exp \left( \frac{\Delta t - \Delta t_0}{\tau} \right),
\end{equation}
\end{linenomath*}

with the resulting values of $\tau$ indicated in the legend. The calculated distributions seem consistent with previous observations \citep[e.g.][]{bieberetal2004}, where the low $\Delta t$ distribution is formed by evaporated neutrons produced in the lead producer. These neutrons have a high level of multiplicity, i.e., every incident high-energy proton leads to the formation of a number of neutrons. The high $\Delta t$ distribution is formed by a combination of low energy neutrons created in the atmosphere, and low-multiplicity neutrons created in the lead producer by low energy protons. \\

The results for the neutron test source (right panel) shows only a single Poisson distribution, as these low energy neutrons probably interact directly with the $^3$He gas, {and do not undergo nuclear reactions with the producer.}\\

\subsection{High-voltage plateau}

{We run the mini-NM tube in the proportionality mode and set the voltage near the observed (and recommended) plateau-region \citep[see, e.g.,][]{kamal2014}.} Although the different tubes' recommended operating voltage regime is given in Table \ref{Tab:tubes}, we have also tested the response of each tube's count rate to voltage changes to characterize any possible changes in the count rate due to changing voltages, and to make sure the operating voltages are, in fact, within the HV plateau for each tube. As a default setting, we operate the tubes at the following voltages: BF$_3$ @ 2300 V, BF$_3$ (thin) @ 1100 V, and $^3$He @ 1350 V. However, with the HV system discussed above, it is easy to adjust this value with simply changing a variable in the corresponding software script. \\

The results of the changing HV values are shown in Figure \ref{fig:HV_testing}. For each tube the count rate was taken for five minutes at each voltage, and where the tests were still performed in the presence of the neutron source. The top panel shows the results for the thin BF$_3$ tube, and these indicate that, for this tube and in this voltage range, there is virtually no response in the count rate. A linear regression is performed (red line), resulting in a gradient of $< 0.001$ \%/V. Similar analyses for the $^3$He (middle panel) and thick BF$_3$ (bottom panel) tubes give gradients of $0.01$ \%/V and $0.03$ \%/V, respectively. Keeping in mind that the HV typically varies less than 1 -- 3 V/min, any errors from a slowly changing HV are indeed negligible within the voltage range shown here.\\

\begin{figure*}
\begin{center}
\includegraphics[width=0.99\textwidth]{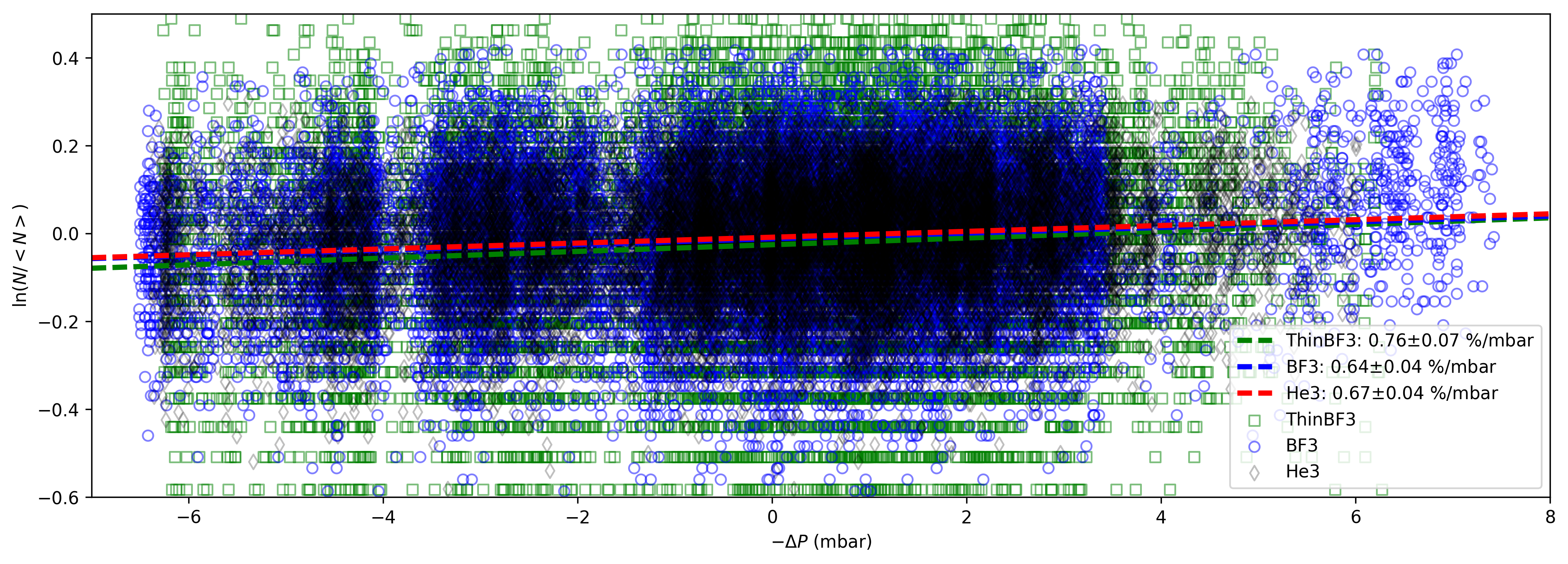}
\includegraphics[width=0.99\textwidth]{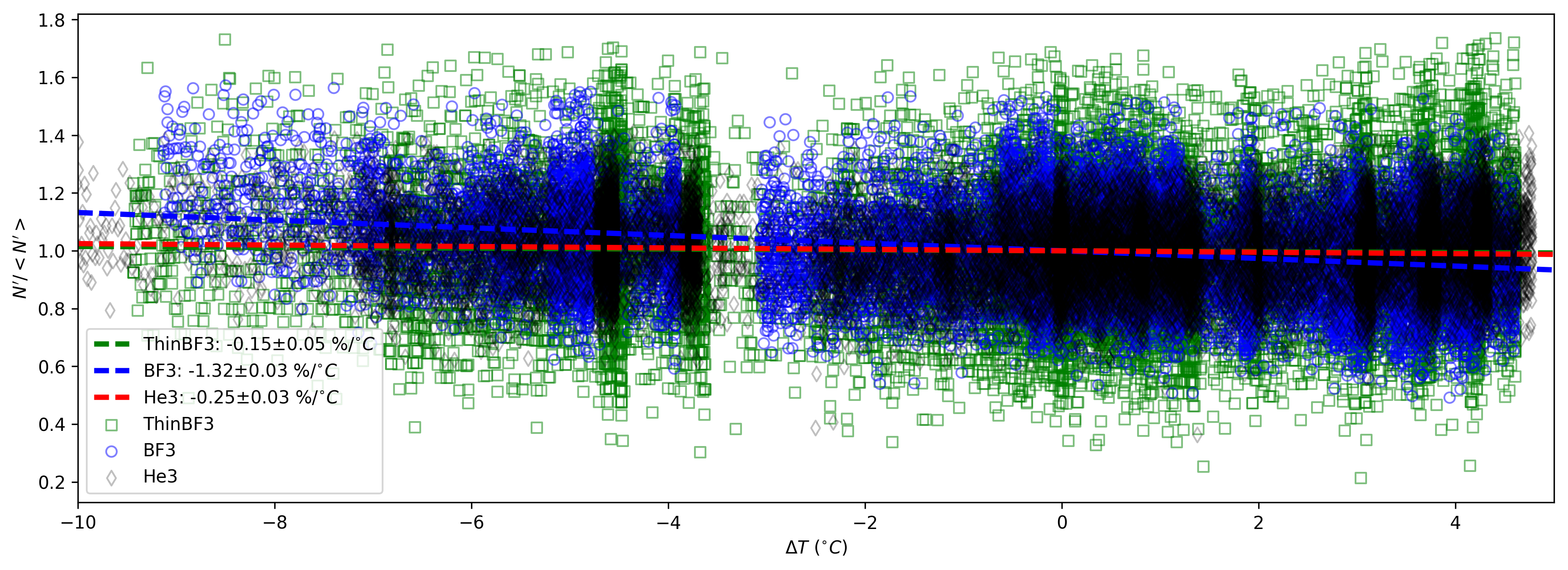}
\caption{The top panel shows the calculation of the barometric coefficient (the values indicated in the legend) for the different tubes (the thin BF$_3$ indicated by green symbols and lines, the BF$_3$ tube by blue, and the results for the $^3$He tube by red). The bottom panel is similar, but now shows a test for any temperature dependence. }
\label{fig:pressure_temp_variations}
\end{center}
\end{figure*}

\begin{figure*}
\begin{center}
\includegraphics[width=0.99\textwidth]{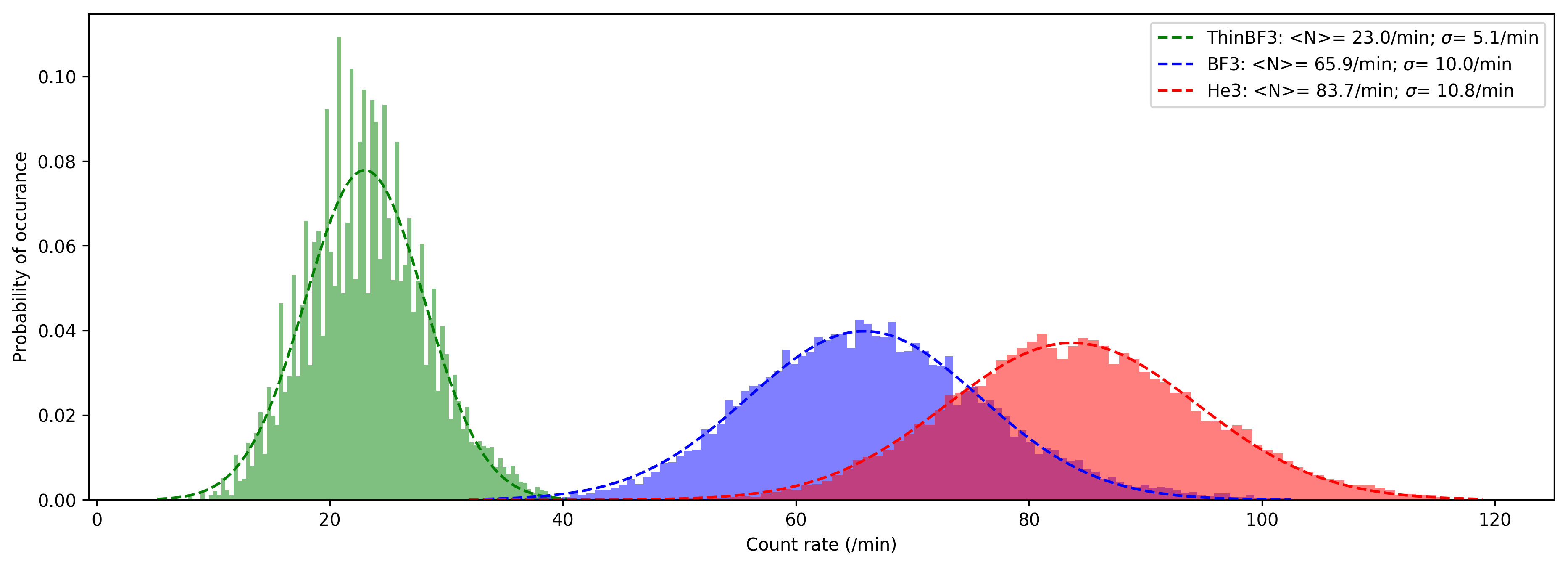}
\caption{Probability distribution of the count rate for each tube (histograms), fitted with a Gaussian curve (dashed lines). The legend shows the average and standard deviation of each fit.}
\label{fig:compare_counts}
\end{center}
\end{figure*}

\section{Cosmic ray measurements: Laboratory testing}

After testing with the neutron source, all three tubes were operated over December 2019 in the physics laboratory to test the pressure and temperature dependence of each tube. We also compare the count rates of the different tubes, this time using cosmic ray neutron measurements over a time interval where the cosmic ray levels were approximately constant. The tests were done in Potchefstroom, South Africa, at a cutoff rigidity of $P_c \sim 7.2$ GV ($26^{\circ}$ 43' S, $27^{\circ}$ 06' E, 1335 m.a.s.l.). \\

\subsection{Pressure correction}

The primary correction that needs to be performed on the count rate is a pressure correction due to the changing atmospheric depth above the monitor. The measured count rate, $N$, changes according to the following relation,

\begin{equation}
    N = \langle N \rangle \exp \left[- \beta \left( P - \langle P \rangle \right)    \right],
\end{equation}

where $P$ indicates pressure, $\beta$ is the barometric coefficient of each tube, and $\langle P \rangle$ indicates the reference pressure (average over a long-time interval). Note that $-\Delta P = - (P - \langle P \rangle )$. More details can be found in e.g. \citet{carmichael1968}, \citet{Paschalisetal2013}, \citet{lararogelio2016}, and \citet{Butikofer2018}.\\

To calculate the barometric coefficient of each tube, we plot $\ln (N / \langle N \rangle)$ as a function of $- \Delta P$, and perform a linear regression, with the barometric coefficient being the gradient thereof. This is shown in the top panel of Figure \ref{fig:pressure_temp_variations} for the different tubes. The calculated barometric coefficients are indicated in the legend and are all within the range $0.65$ -- $0.75$ \%/mbar. Once these coefficients are known, we apply the pressure correction by

\begin{equation}
    N' = N \exp \left(  \beta \Delta P \right),
\end{equation}

where $N'$ is the pressure corrected count rate of each tube.

\subsection{Temperature dependence}

Previous studies \citep[e.g.][]{evensonetal2005,krugeretal2008} have shown that NM count rates are affected by temperature variations. It should be noted that, in contrast to the real barometric effect, related to the physics of atmospheric cascade, the temperature effect is spurious and related to the stability of the detector rather than to the real changes of the neutron flux. In traditional NMs, the temperature is usually carefully maintained, and these effects are not important. However, the more mobile mini-NMs are often placed in environments with large temperature changes, and hence, we need to check for (and correct if necessary) any possible systematic temperature effects. Any temperature effects are hard to predict, and different components of the mini-NM probably react differently to temperature variations \citep[see the discussion by][]{evensonetal2005}.\\

The bottom panel of Figure \ref{fig:pressure_temp_variations} shows $N'/ \langle N' \rangle$ (note this is already the pressure-corrected count rate) as a function of the temperature change $\Delta T = T - \langle T  \rangle$ for the different tubes. Note that we use the internal temperature sensor that is placed in-between the tube and the moderator. All three tubes show a negative temperature dependence, with the count rate decreasing with increasing temperature. For the thin BF$_3$ tube, the temperature dependence is, however, negligible, while it may become significant for the other tubes, especially if these are placed in an environment with large temperature changes. The correction for the temperature changes can be implemented as

\begin{equation}
    N_T' = N' - \beta_T \Delta T,
\end{equation}

where $N'_T$ is now the temperature and pressure corrected count rate, and $\beta_T$ the so-called {\it temperature coefficient} (i.e., the gradient of the linear regression curve). Both linear regressions discussed above go, by definition, through the origin. However, depending on the mini-NM set-up and its dependence on the local environmental conditions, a linear regression may not be sufficient, and higher-order polynomials might be required. Some earlier works have also implemented exponential functions \citep[e.g.,][]{heber2015}, which can be reduced to a linear approximation if $\beta_T \Delta T$ is sufficiently small. \\

Similar to the barometric coefficient, we should note that the temperature coefficient might be highly dependent on the local environmental conditions where the monitor is placed and that the pressure and/or temperature coefficients given here cannot be universally applied \citep[see, e.g.,][]{krugermoraal2010,iemsaad2015}. Generally we do not test for any humidity or atmospheric water vapour effects \citep[see][]{Banglieng2020}.\\

\subsection{Comparison between the different tubes}

The pressure corrected count rate from each tube is binned as a histogram, and shown in Figure \ref{fig:compare_counts}. Gaussian distributions are fitted to each histogram (dashed lines), showing that the remaining variations are roughly consistent with statistical cosmic ray variations. The average count rate (as derived from the fitted distribution) from each tube is indicated in the legend, along with the standard deviations. The efficiency of the different tubes are clearly illustrated in this figure: The thin BF$_3$ tube is, as expected, the least efficient with an average count rate of 23 counts/min, followed by the BF$_3$ tube with 66 counts/min, and the $^3$He tube being the most efficient with a count rate of 84 counts/min. However, the increasing efficiency comes with a price, with the $^3$He being, by far, the most expensive, and the thin BF$_3$ being, by far, the most inexpensive of the tubes tested. As such, the BF$_3$ tube is recommended for a standard mini-NM set-up, balancing statistical accuracy with operating costs. It should be noted that the tube used depends on the application, i.e., if one is interested in low resolution data (i.e., data or weekly averages for soil moisture measurements), then a thin BF$_3$ tube could still be sufficient for the task at hand. \\

In addition, a neutron monitor is often considered as an instrument for long-time measurements. Some cosmic ray stations have neutron monitors in operation for 50 years or even longer. {In such applications, BF$_3$ tubes show very good stability while $^3$He-filled ones can experience a long-term drift of parameters because of leakage of helium gas. On the other hand, the efficiency of BF$_3$ counters may also gradually decline because of deposition on the anode wire that can become notable for high-latitude, high-elevation NMs with very high count rates, such as South Pole \citep[][]{Bieberetal2007}.}




\begin{table*}
\caption{Current and planned mini-neutron monitor stations.}             
\label{Tab:mini_NM_network}
\centering                          
\begin{tabular}{c c c c c c}        
\hline\hline
{\bf Station name} & {\bf Lat. ($^{\circ}$)}  & {\bf Long. ($^{\circ}$)} & {\bf Alt. (m.a.s.l.)} & {\bf Tube} & {\bf Comments} \\ 
\hline  
{\bf Current stations} & & & & & \\
\hline
Dome C (Antarctica) & 75$^\circ$06'S & 123$^\circ$23'E & 3233 & BF3 & Lead and lead-free unit \\
Sierra Negra Peak (Mexico) & 18$^{\circ}$59'N & 97$^{\circ}$19'W & 4100 & BF3 & Uses older electronics \\
Riyadh (Saudi Arabia) & 24$^{\circ}$38'N & 46$^{\circ}$43'E  & 612 & BF3 &  \\
Zugspitze (Germany) & 47$^{\circ}$25'N  & 10$^{\circ}$58'E & 2650 & BF3 &  \\
\hline
{\bf Future/planned stations} & & & & & \\
\hline
Addis Ababa (Ethiopia) & 9$^{\circ}$06'N & 38$^\circ$46'E & 3293 & BF3 & Deployment in 2021 \\
Chacaltaya (Bolivia) & 16$^{\circ}$21'S  & 68$^{\circ}$07'W & 5240 & BF3 & Deployment in 2021 \\
\hline
{\bf Mobile stations} & & & & & \\
\hline
SA Agulhas II (South Africa) & $\sim$33$^{\circ}$S -- 72$^{\circ}$S & $\sim$ 37$^{\circ}$W -- 37$^{\circ}$E & $\sim 0$ & He3 &  \\
Polarstern (Germany) & $\sim$85$^{\circ}$N -- 70$^{\circ}$S & $\sim$ 70$^{\circ}$W -- 20$^{\circ}$E & $\sim 0$ & He3 & Uses older electronics \\
BIO Hesp{\'e}rides (Spain) & $\sim$37$^{\circ}$N -- 62$^{\circ}$S  & $\sim$ 60$^{\circ}$W -- 0$^{\circ}$E & $\sim 0$  & BF3 &  \\
\hline
\hline
\end{tabular}
\end{table*}

\section{Discussion}

Below we briefly describe measurements from three mini-NM stations producing long-term data, while Table \ref{Tab:mini_NM_network} summarizes the current and planned mini-NM stations.

\subsection{Concordia research station}

A pair of mini-NMs was installed at the French-Italian research station Concordia (Dome C, Antarctic plateau) at 75$^\circ$06' S, 123$^\circ$23' E, 3233 m a.s.l. in 2015 \citep[][]{stepan2015,usoskin2015}. One of those instruments (called DOMC) is a standard-design unit with the lead producer and the the other is ``bare" (lead-free, DOMB). Both use the thick BF$_3$ tube discussed here. The unique location of the station (high elevation and close proximity to the geomagnetic pole) provides low atmospheric and negligible geomagnetic cutoffs. Moreover, DOMC and DOMB are {ones of few} NMs whose asymptotic cone point south of the equatorial plane, making it possible to study the 3D distribution of cosmic-ray variability. The average count rates of DOMC and DOMB are about 1150 and 315 counts/min, respectively, during 2019. The high altitude polar location makes these mini-NMs exceptionally sensitive to low-energy cosmic rays, which is particularly important for registration of solar energetic particles (SEPs). In confirmation of that, several SEP events have been registered since the start of the operation of mini-NMs at Concordia, including very weak ones \citep[e.g.,][]{Mishev_JSWSC2017}. {For example, GLE \#72 on 10 September 2017 caused count rate increases of +10\% (maximum 1210 cts/min with pre-increase 1102 cts/min) and +13\% (maximum 330 cts/min with pre-increase 293 cts/min) at DOMC and DOMB, respectively  (see the International GLE database \url{http://gle.oulu.fi}). Introduction of those instruments} to the neutron monitor global network significantly increased its overall sensitivity to SEPs and led to a revision of the ground level enhancement (GLE) definition with the introduction of the sub-GLE class of events \citep[][]{poluianov_SP2017}. In 2019, neutron monitors DOMC and DOMB were upgraded with the electronics described in this paper (in November and August, respectively). Before that, their electronics were similar to one in \citet{Krueger2013}.\\

\subsection{Neumayer III}

A mini-NM (using a thick BF$_3$ tube) was installed at the German Antarctic Neumayer III research station early in 2011 \citep[][]{heber2015}. Neumayer III is located on the Ekstr{\"o}m Ice-shelf in the Atka-bay in the Weddell-sea at 70$^\circ$~40~S, 8$^\circ$16~W. The corresponding cutoff rigidity $P_c=0.1$~GV has been calculated utilizing PLANETOCOSMICS \citep{Desorgher-etal-2009} with the International Geomagnetic Reference Field \citep[][]{Finlayetal2010} and the \citet{Tsyganenko1989} as model for the inner and outer Earth magnetosphere, respectively \citep[][]{Herbstetal2013}. The monitor provided data from 2013 -- 2017, and was primarily used to study the long-term stability of the instrument as there were few transient cosmic ray events during this period. The monitor was transport back to Kiel, Germany, in 2018, and retro-fitted with the new electronics discussed in this paper in early 2020. It is currently undergoing testing, and will be redeployed in the near future.\\

\subsection{Polarstern}

In 2011, a mini-NM with a $^3$He tube, was installed on the German research vessel, the Polarstern \citep[][]{heber2015}. The instrument has provided data since 2012, and is used to study both the long-term stability and sensitivity of the instrument, as well as the latitude effect as the ship sails each year from Germany to Antarctica.\\

\subsection{Sierra Negra}

This mini-NM was deployed on a plateau at the slope of the Sierra Negra volcano in the central part of Mexico (19.0$^{\circ}$N, 97.3$^{\circ}$W), where the High Altitude Water Cherenkov (HAWC) gamma ray observatory is located. The cutoff rigidity at this site is $\sim 7.8$ GV and the altitude is 4100 m.a.s.l., allowing relatively high count rates of $\sim 840$ counts/min (on January 2018). The  mini-NM has been taking data since 2015, with a data gap of three months during 2016. Taking advantage of the portability of the mini-NM, an altitude survey was performed during 2014 going from the sea level up to 4580 m.a.s.l. \citep[][]{lararogelio2016}. This study showed that the mini-NM is stable and performs well in the field.\\

\section{Summary and outlook}

The original purpose of the mini-NM was to develop a small and portable NM that can be used to inter-calibrate the world-wide network of NMs. These stationary monitors, placed at different altitudes and magnetic latitudes, also have different environmental conditions, and react to environmental changes differently. We believe that such an inter-calibration study could still be relevant, and should, at some point, be continued. \\

On the other hand, mini-NMs stations have produced valuable data and scientific insights, especially at polar regions and high altitudes where the mini-NM count rates are comparable to full-scale sea-level NMs. We firmly believe that this trend will continue, and, in the future, a world-wide network of mini-NMs will co-exist with the current network of NMs, especially covering current observational gaps (e.g., the sparse coverage over the Southern Hemisphere including African and South America). The mini-NMs are much more cost-effective to construct and operate than traditional NMs, providing an intriguing possibility to also retrofit older stations that were decommission in the past due to financial constraints with low maintenance mini-NMs. Most of these sites already have the necessary infrastructure (including power and, in some cases, internet access). Potential sites include ‘Sulphur Mountain, Canada, at 2383 m, the site of the original NM64 monitor.\\

\acknowledgements

{Operation of DOMC/DOMB mini-NMs is supported by projects CRIPA and CRIPA-X (No. 304435) of the Academy of Finland, Finnish Antarctic Research Program (FINNARP). DOMC/B NM data can be obtained from \url{http://cosmicrays.oulu.fi}, courtesy of the Sodankyl\"a Geophysical Observatory. Operation of DOMB/DOMC NMs are possible thanks to the hospitality of the Italian polar programme PNRA (via the LTCPAA PNRA 2015/AC3 project) and the French polar Institute IPEV. The neutron monitor database NMDB  (\url{http://nmdb.eu}) founded under the European Union's FP7 programme (contract no. 213007) is acknowledged for hosting and providing data. We acknowledge support from the MINECO - FPI 2017 program, co-financed by the European Social Fund. This work has been supported by the project CTM2016-77325-C2-1-P, funded by Ministerio de Econom\'ia y Competitividad, and by the European Regional Development Fund, FEDER. This work is based on the research supported in part by the National Research Foundation of South Africa (NRF grant number: 119424). Opinions expressed and conclusions arrived at are those of the authors and are not necessarily to be attributed to the NRF. Figures prepared with Matplotlib \citep[][]{Matplotlib-2007}.}


\bibliographystyle{swsc}
\bibliography{biblio}

\end{document}